\documentclass[twocolumn,superscriptaddress,amsmath,amssymb,aps,prd,longbibliography,10pt]{revtex4-2}
\usepackage{dcolumn}
\usepackage{bm}
\usepackage{graphicx}
\usepackage{color}
\usepackage{comment}

\usepackage{amsthm}

\newcommand{\be}{\begin{equation}}
	\newcommand{\ee}{\end{equation}}
\newcommand{\ba}{\begin{eqnarray}}
	\newcommand{\ea}{\end{eqnarray}}

\newcommand{\n}[1]{\label{#1}}
\newcommand{\eq}[1]{(\ref{#1})}

\newcommand{\hhh}{\, ,\hspace{0.5cm}}

\newcommand{\ind}[1]{\mbox{\tiny #1}}

\newcommand{\mpla}{m_{\ind{P}}}

\begin{document}

%\title{On $f(R)$-gravity off-shell equivalence}

\title{On the equivalence between $ f\left(R\right) $ theories and Einstein gravity}

\author{Soham Bhattacharyya}
\email{xeonese@gmail.com}
%\author{Andrey A. Shoom}
%\email{andrey.shoom@aei.mpg.de}
\affiliation{Department of Physics, IIT Madras, Chennai, India}
	
%\date{\today}
	
\begin{abstract}
In this brief note we present a somewhat trivial result. Namely, we show that perturbative off-shell $f(R)$-theory is equivalent to Einstein gravity, as well as to the Brans-Dicke theory and the Einstein scalar field model. We also discuss possible generalisation of this result to higher-order gravitational field models.
\end{abstract}
	
\maketitle
	
\section{Introduction}
	\begin{comment}
{\bf LAYOUT: the first sketch of}
\begin{itemize}
	\item Field quantisation: main idea, QED success. Should gravitational field be quantised? (very briefly)
	
	Success in quantisation of electromagnetic field followed by the qunatization of electroweak interaction 
	
	\item Approaches taken to quantise gravity: ``canonical way", supersymmetry, string theory, loop quantum gravity (give only the key ideas and problems faced)
	\item The low energy limit: higher-order Lagrangian (to formulate the problem)    
	\item The related models and results: what people did and where do they go
	\item Emerged models: Gauss-Bonet terms, higher-order curvature therms, $f(R)$-model
	\item f(R) in classical gravity and cosmology: a brief review
	\item Our goals: what do we do here, why we do that, what we got, what it means and what it gives
	\item Organisation of our work
	\item Conventions and units 
\end{itemize}
	\end{comment}
Significant strides were made in the early parts of the twentieth century in the formulation of gravity as a classical field theory. Out of that effort came forth General Relativity (GR), which passed almost all the tests thrown at it, starting from the short, terrestrial (or weak gravity region) length scales \cite{Kapner:2006si}, to solar system tests \cite{Will:2014kxa}, to binary pulsar (strong gravity region) tests \cite{Will:2014kxa}, and finally constraints from Gravitational Wave (GW) (extreme gravity region) observations in \cite{Ghosh:2017gfp}.\\

GR is a theory with two massless degrees of freedom. The degrees of freedom manifest at asymptotic infinity as the so called plus and cross polarizations in gravitational wave detectors. Efforts have been made to ascertain whether there are more than two degrees of freedom in the observed gravitational wave signal, as in \cite{LIGOScientific:2016lio,LIGOScientific:2021sio,Krishnendu:2021fga}, and others. However, so far there is no conclusive proof of the presence of more than two degrees of freedom. On the other hand, there are more than enough theoretical and observational considerations that GR cannot be a complete theory of gravity. In the theoretical side, prediction of singularities in extreme situations like black holes or big bang render the theory useless in such scenarios. Similarly, in the observational side, the presence of flat velocity profile of galaxies instead of decaying ones, or the observational challenges faced by the $ \Lambda $CDM model provide much needed motivation to come up with modified theories of gravity.\\

However, modifying gravity has its costs. Most modifications to gravity leads to certain observational features that have so far eluded experiments. Generally speaking, modifications to GR leads to either increase in degrees of freedom or violations of one or more of the postulates of GR. On the increase in degrees of freedom aspect, it is possible for gravity to have six degrees of freedom, given that we live in a four dimensional space-time. In the most naive sense, one can make coefficients of the modification to GR go to very small values in the action level and claim that the modifications are un-excitable. However, that theory loses any predictive power. However, there are other ways in which one can show modifications to gravity can be ``pushed" to sub-dominant regimes. For example, in some special cases it can be shown that the extra degrees of freedom turn out to be non-propagating (hanging around perturbed black holes as some massive scalar cloud), like for the case of Ricci flat solutions of $ f(R) $ gravity (see Chapter 6 of the thesis \cite{Bhattacharyya:2019zna}), or it is shielded by means of the so called \textit{Chameleon} mechanism \cite{Hees:2013xna}. Such a mechanisms dictate that extra degrees of freedom become highly massive (that is they cannot be excited by standard solar system excitations) in dense environments and light in very low density scenarios, leading to expansion of the universe as a whole. Similarly, in the violations of postulates aspect, for example, in Chern-Simons theory of gravity, a pseudo-scalar fields breaks the parity invariance of GR and leads to the birefringence of vacuum, as in \cite{Jackiw:2003pm}.\\

There exists another little known method to ``push" leading order modifications to GR to sub-leading orders. This is the case of the field redefinition. In such cases, the metric is redefined in a particular manner such that leading order effects (such as quadratic modifications to GR) can be pushed via conformal transformations to cubic orders or higher. For example, in \cite{deRham:2020ejn}, it was shown that under field redefinition of the metric tensor, certain combinations of the curvature scalars appearing as modifications to the Einstein-Hilbert action can be pushed to higher orders. That is, any observer whose effective metric is the redefined metric will no longer \textit{see} any quadratic modification, as well as some cubic modifications. Since the energy scales at which quadratic modifications appear are lower than the energy scales of the cubic modifications, it becomes harder for any astrophysical source to perturb or excite the additional degrees of freedom in the redefined metric scenario. In the language of \cite{deRham:2020ejn}, dimension four terms have been pushed to dimension eight under the field redefinition. In this article, we generalize the quadratic order calculations of \cite{deRham:2020ejn} to arbitrary powers of the Ricci scalar.\\

At the low energy limit, string theoretic/loop quantum gravity or in other words, effective field theory actions can lead to a Lagrangian of the following form
\begin{eqnarray}\label{1.1}
S &=& \frac{1}{16\,\pi}\,\int\,d^4x\,\left[R\,+\,\mathcal{R}^2\,+\,\mathcal{R}^3 \,+\, \mathcal{O}\left(\mathcal{R}^4\right) \,+\,\mathbf{\mathcal{L}_m} \right],\nonumber\\
\end{eqnarray}
where $ \mathcal{R}^n $ are general curvature invariants of dimension $ 2\,n $ consisting of the Ricci scalar, \textcolor{black}{Ricci tensor}, and the Riemann tensor. $ \mathcal{L}_m $ is the matter Lagrangian with a minimal coupling to the curvature terms, and is also a functional of the metric tensor. If in Eq. (\ref{1.1}), one only considers till curvature squared terms of the most general form like that of the Gauss-Bonnet type,
\begin{eqnarray}\label{1.2}
\mathcal{R}^2 &=& a\,R^2 \,+\, b\,R_{\mu\nu}\,R^{\mu\nu} \,+\, c\,R_{\mu\rho\nu\sigma}\,R^{\mu\rho\nu\sigma},
\end{eqnarray}
then it was shown by \cite{Stelle:1976gc} that such a model is renormalizable.\\
	
Authors like \cite{Slovick:2013rya} decided to approach the inverse problem while trying to perturbatively renormalize gravity. They started with GR, and under field redefinition was able to obtain the most general quadratic theory, that is Eq. (\ref{1.1}) truncated till $ \mathcal{R}^2 $.\\
	
In this article we do not pursue all possible type of independent scalar invariants at each dimension, but restrict ourselves to only vacuum polynomial $ f\left(R\right) $ theories of the form
\begin{eqnarray}\label{1.3}
S=\frac{1}{16\pi}\int d ^4x\sqrt{-g}\left[R+f(R)\right]\,,
\end{eqnarray}
where $ R $ is the Ricci scalar and the form of $ f\left(R\right) $ is given in Eq. (\ref{3.1}). Whereas GR has only two massless spin-2 (or purely tensorial) degrees of freedom (dof), $ f\left(R\right) $ theories of gravity on the other hand contain, in addition to the massless dof of GR, an extra massive scalar degree of freedom. This is best understood by going to the so called `Einstein frame' by means of a conformal transformation of the original frame (or `Jordan frame') metric of the $ f\left(R\right) $ theory which frames the theory effectively as GR plus scalar field with a potential (the form of which is fixed by the particular form of the $ f\left(R\right) $ theory at hand). Examples of such can be found in \cite{Shtanov:2022wpr,RevModPhys.82.451,DeFelice:2010aj,WHITT1984176,BARROW1988515,Sotiriou_2006}.\\
	
We however find that in perturbative $ f\left(R\right) $ theories of gravity, with $ f\left(R\right) $ of the form of Eq. (\ref{3.1}), under a suitable field redefinition, can be mapped back to GR \textit{ad infinitum.} We find that the field redefinition responsible for mapping the theory back to GR is a conformal transformation on the metric.\\
	%conformal \cite{Shtanov:2022wpr}
	%reviews \cite{RevModPhys.82.451,DeFelice:2010aj}
	%equivalence \cite{WHITT1984176,BARROW1988515,Sotiriou_2006}
	
Our paper is organized as follows. In Sec.~II we briefly review the Einstein and $f(R)$-gravity theories. In Sec. III we review the field redefinition method to excise quadratic (and some cubic) modifications to GR. In Sec.~IV we prove the perturbative off-shell equivalence of these theories. Section V contains a brief review of Brans-Dicke theory and describe its off-shell equivalence with $f(R)$-theory. In Sec.~VI we review off-shell equivalence between $f(R)$-gravity and the Einstein-scalar field model. In Sec.~VII we establish perturbative off-shell equivalence between $f(R)$-gravity, Brans-Dicke theory, and the Einstein-scalar field model and illustrate it via equivalence maps. Finally, in Sec.~VIII we conjecture on the potential effects of the conformal transformation when matter fields are present.\\
	
In this paper we shall use the geometrized system of units, setting $c=G=1$. We use space-time signature $+2$ and conventions adopted in the book \cite{MTW}, unless explicitly stated otherwise. %We shall express all physical quantities in units of the Planck mass $\mpla=\sqrt{1/G}\approx10^{19}$GeV. 
	
\section{Einstein and $f(R)$-gravity models}
	
The famous 4-dimensional Einstein-Hilbert action in the presence of matter is
\be\n{2.1}
S_{\ind{EH}}=\frac{1}{16\pi}\int d^4x\sqrt{-g}R + S_{m}(g_{\alpha\beta},\psi)\,,
\ee
where $g$ is determinant of space-time metric $g_{\alpha\beta}$ in the chosen coordinates $x^{\alpha}$, $R$ is the Ricci scalar, and the integral is taken over the space-time manifold. The matter action $S_{m}$ contains matter fields $\psi$. One of the simplest generalization of the Einstein-Hilbert action is the so-called $f(R)$ model,
\be\n{2.2}
S=\frac{1}{16\pi}\int d ^4x\sqrt{-g}\left[R+f(R)+\mathbf{\mathcal{L}_m}\right]\,,
\ee
where $f(R)$ is an arbitrary function of the Ricci scalar. %{\bf say more about these models; equivalence principle}\\
	
$ f\left(R\right) $ theories shot to prominence with the quadratic version of Eq. (\ref{2.2}), or the `Starobinsky model' given by $ R\,+\,\alpha\,R^2 $ (\cite{Starobinsky:1980te}), which was shown to demonstrate inflation, and remains a viable alternative to scalar field inflationary models.\\ %The model's built in features contain the observed temperature anomalies of the CMB as well.\\
	
Equations of motion or field equations which have higher than second order derivatives suffer from an instability where their total energy or the Hamiltonian is unbounded from below. Implying that in general these theories do not have a stable vacuum state. This is referred to as the Ostrogradski instability in literature. $ f\left(R\right) $ theories of gravity however do not suffer from the Ostrogradski instability, as was found in \cite{Woodard:2006nt}. This is in contrast to more generalized theories of gravity containing scalar invariants like Ricci tensor and Riemann contractions whose field equations are also fourth order in nature, and they suffer from the so called `ghost' degrees of freedom \cite{Stelle:1976gc} whose energy is unbounded from below.\\
	
There are two different formalisms of $ f\left(R\right) $ theories of gravity, the metric and the Palatini (or metric-affine) formalism. While the metric formalism is the standard that assumes the metric to be the dynamical field in the space-time, the Palatini formalism takes the connection to be an independent variable as well in addition to the metric tensor (a review in \cite{Olmo:2011uz}), leading to some interesting features, like the suppression of the massive scalar degree of freedom as was found in \cite{Olmo:2019flu}.\\
	
It has been noticed by more and more accurate experiments that there exist in nature, an equivalence (or a strict equality) between the gravitational `charge' (or gravitational mass) and the inertial mass. GR as a theory satisfies the equivalence principle. The fate of the equivalence principle in $ f\left(R\right) $ theories of gravity is still debated in the gravity community. On one hand, various authors argue that the principle is violated, for example in the Palatini formalism in \cite{Olmo:2006zu}. Also, since $ f\left(R\right) $ theories can be mapped to scalar-tensor theories of a particular form (as in \cite{Sotiriou:2006hs}), it was shown in \cite{Blasone:2018ftv} that equivalence principle is violated in such theories. On the other hand, it was shown in \cite{Khoury:2003rn} that for transition from dense regions of space-time to vacuum, the exterior solution can be represented by a Schwarzschild-like metric, with the extra Yukawa type field (arising due to $ f\left(R\right) $ modification) only appearing as a very thin shell around the dense object. In such a case, the weak equivalence principle is shown to hold to a very high accuracy, like for solar system tests of the equivalence principle. Constraints on fifth force parameters can also be found in \cite{Khoury:2003rn}

\section{The field redefinition method}

Following the notations of \cite{deRham:2020ejn} only for this section, we briefly review the field redefinition method here. For a Lagrangian of the following form \cite{TSEYTLIN198692} in vacuum
\begin{eqnarray}\label{lagrangian}
	\mathcal{L} &=& \sqrt{-g}\,\frac{M_{Pl}^2}{2}\,R + \mathcal{L}_{D4} + \mathcal{L}_{D6} + \mathcal{O}\left(\frac{Riemann^4}{M^4}\right),\nonumber\\
\end{eqnarray}
where the first term is the Einstein-Hilbert action and the higher dimensional operators are
\begin{eqnarray}
	\mathcal{L}_{D4} &=& \sqrt{-g}\,\left[c_{R^2}\,R^2+c_{W^2}\, W^2_{\mu\alpha\nu\beta}+c_{GB}\,R^2_{GB}\right], \\
	\mathcal{L}_{D6} &=& \frac{\sqrt{-g}}{M^2}\,\left[d_1R\Box R+d_2R_{\mu\nu}\Box R^{\mu\nu}+d_3R^3\right.\nonumber\\
	&&d_4 R R^2_{\mu\nu}+d_5 R R^2_{\mu\nu\alpha\beta}+d_6 R^3_{\mu\nu}+d_7R^{\mu\nu} R^{\alpha\beta} R_{\mu\nu\alpha\beta}\nonumber\\
	&&d_8 R^{\mu\nu} R_{\mu\alpha\beta\gamma} R_\nu^{\;\;\alpha\beta\gamma}+d_9 R_{\mu\nu}^{\;\;\;\;\alpha\beta} R_{\alpha\beta}^{\;\;\;\;\gamma\sigma} R_{\gamma\sigma}^{\;\;\;\;\mu\nu}\nonumber\\
	&&\left.+d_{10} R_{\mu\;\;\;\nu}^{\;\;\alpha\;\;\;\beta} R_{\alpha\;\;\;\beta}^{\;\;\gamma\;\;\;\sigma} R_{\gamma\;\;\;\sigma}^{\;\;\mu\;\;\;\nu}\right],
\end{eqnarray}
where $ W_{\mu\alpha\nu\beta} $ is the Weyl tensor and $ R^2_{GB} $ is the Gauss-Bonnet topological term which can be ignored since in this article we restrict ourselves to four dimensions. One can perform a perturbative field redefinition of the metric as follows
\begin{eqnarray}
	g_{\mu\nu}&&\rightarrow g_{\mu\nu}-\frac{2}{M_{Pl}^2}\left[-2c_{W^2} R_{\mu\nu}+\left(c_{R^2}+\frac{1}{3}c_{W^2}\right)g_{\mu\nu}R\right]\nonumber\\
	&&-\frac{2}{M_{Pl}^2}\frac{1}{M^2}\left\{-d_2\Box R_{\mu\nu}-d_4 R R_{\mu\nu}-d_6 R_\mu^{\;\;\alpha} R_{\nu\alpha}\right.\nonumber\\
	&&-d_8 R_\mu^{\;\;\alpha\beta\gamma} R_{\nu\alpha\beta\gamma} + g_{\mu\mu}\left[\left(d_1+\frac{d_2}{2}\right)\Box R+\left(d_3\right.\right.\nonumber\\
	&&\left.\left.\left.+\frac{d_4}{2}\right)R^2+\frac{d_6+d_7}{2}\,R^2_{\alpha\beta}+\left(d_5+\frac{d_8}{2}\right)R^2_{\alpha\beta\gamma\sigma}\right]\right\}, \nonumber\\
\end{eqnarray}
which leads to the Lagrangian (\ref{lagrangian}) becoming as follows
\begin{eqnarray}
	\mathcal{L}&=& \sqrt{-g}\left[\frac{M_{pl}^2}{2}\,R + \frac{1}{M^2} \left(d_9 R_{\mu\nu}^{\;\;\;\;\alpha\beta} R_{\alpha\beta}^{\;\;\;\;\gamma\sigma} R_{\gamma\sigma}^{\;\;\;\;\mu\nu}\right.\right.\nonumber\\
	&&\left.\left.+d_{10} R_{\mu\;\;\;\nu}^{\;\;\alpha\;\;\;\beta} R_{\alpha\;\;\;\beta}^{\;\;\gamma\;\;\;\sigma} R_{\gamma\;\;\;\sigma}^{\;\;\mu\;\;\;\nu}\right)\right]+\mathcal{O}\left(\frac{1}{M^4}\right).
\end{eqnarray}
One can see that due to the field redefinition entirety of the quadratic and some of the cubic terms, except pure Weyl cube terms which cannot be field redefined away, have been pushed to higher orders. In the following section, ignoring Ricci tensor and Riemann tensor terms in the Lagrangian, we generalize the field redefinition method to higher powers of the Ricci scalar.
		
\section{Equivalence with Einstein gravity}
	
Here we shall consider perturbative approach to $f(R)$ gravity. Namely, we introduce a parameter $\lambda$ which characterizes strength of perturbation, such that $\lambda R\ll1$. This parameter can be considered as squared characteristic length of the model (measured in units of $\mpla$). In the string effective action it corresponds to the string slope parameter $\alpha'$, which in natural units is equal to the squared string length. Hence, the dimension of $ \lambda $ is length squared. The perturbative $f(R)$ takes the following form (see, e.g. \cite{METSAEV198752,deRham:2020ejn}):
\be\n{3.1}
f(R)=\sum_{k\geq1}c_{k}\lambda^{k}R^{k+1}\,,
\ee
where $c_{k}$'s are dimensionless expansion coefficients \footnote{For the given choice of the perturbation parameter $\lambda$ the integrand in \eq{2.2} reads $R(1+\lambda R +\lambda^{2}R^{2}+\cdots)$.  However, one could also consider $\lambda$ as the characteristic length, such that $\lambda^{2}R\ll1$, and have $R(1+\lambda^{2}R +\lambda^{3}R^{2}+\cdots)$. One can establish a transformation between these choices.}. Our goal is to establish perturbative equivalence between the $f(R)$ theory \eq{2.2} with $f(R)$ given by \eq{3.1} and the Einstein gravity \eq{2.1}. Such an equivalence was established between Lagrangian density with quadratic and cubic terms in $R$, $R_{\alpha\beta}$, and the Riemann tensor components in earlier works by means of field redefinition (see, for example, \cite{TSEYTLIN198692,METSAEV198752,BARROW1988515,AccettulliHuber:2019jqo}). In the case of $R$ terms only, the field redefinition is equivalent to conformal transformation of space-time metric. Here we shall use the same approach, which can also be suggested by the Scherk-Schwarz formalism for performing dimensional reduction, which we do not apply here \cite{Scherk:1979zr}. 
	
The conformal metric transformation
\be\n{3.2}
\bar{g}_{\alpha\beta} = \Omega^2\,g_{\alpha\beta}\,,
\ee
induces the following \textcolor{black}{Ricci scalar} conformal transformation in 4-dimensional space-time (see e.g. \cite{Wald}):
\be\n{3.3}
\bar{R}=\Omega^{-2}\left(R-6\,\Omega^{-1}\nabla^{2}\Omega\right)\,,
\ee
where $\nabla$ is the covariant derivative operator associated with the space-time metric $g_{\alpha\beta}$, $\nabla_{\alpha}\,g_{\beta\gamma}=0$. Here and in what follows, we shall use the notations $\nabla^{2}=g^{\alpha\beta}\nabla_{\alpha}\nabla_{\beta}$ and $(\nabla\Omega)^{2}=g^{\alpha\beta}(\nabla_{\alpha}\Omega)(\nabla_{\beta}\Omega)$. It is to be noted that under the conformal transformation (\ref{3.2}), the coupling of matter with the space-time curvature changes. This is because of the fact that the matter Lagrangian density $ \mathcal{L}_m $ explicitly contains the metric, and any redefinition of the metric will correspondingly feature in $ \mathcal{L}_m $ as well. The point has been illustrated in \cite{Ruhdorfer:2019qmk}. However, for simplicity, and to illustrate our point, we will only consider the vacuum case.
	
Now we require that for a certain conformal factor the $f(R)$ theory \eq{2.2} is equivalent to the Einstein gravity \eq{2.1}, that is
\be\n{3.4}
\sqrt{-g}\left[R+f(R)\right]\circeq\sqrt{-\bar{g}}\bar{R}\,,
\ee
where $\bar{g}$ is determinant of $\bar{g}_{\alpha\beta}$ and the symbol $\circeq$ stands for equality modulo total derivatives, which give vanishing within our theory boundary terms. Let us remark that expression with $\circeq$ is always understood under the integral sign, i.e. we can multiply the expression like \eq{3.4} by a constant, but not by a scalar function, except for a function of the metric $g_{\alpha\beta}$ and its determinant. Using the expression \eq{3.3} and the relation $\sqrt{-\bar{g}}=\Omega^{4}\sqrt{-g}$ we present \eq{3.4} in the following form:
\be\n{3.5}
6\Omega\nabla^{2}\Omega-(\Omega^{2}-1)R+f(R)\circeq0\,.
\ee   
Let us observe first that for $f(R)=0$ the expression \eq{2.2} reduces to the expression \eq{2.1}, which implies that $\Omega=1$, which is a trivial solution to \eq{3.5} with $f(R)=0$. Because $f(R)$ is given in terms of series \eq{3.1}, we shall look for a solution to \eq{3.5} in the perturbative form of the corresponding to $f(R)$ order,
\be\n{3.6}
\Omega=1+\sum_{k\geq1}\lambda^{k}P_{k}[R]\,,
\ee
where $P_{k}[R]=P_{k}(R,\nabla^{2}R,...)$ are sought polynomials composed of the Ricci scalar and its covariant derivatives. Then, the conformal factor $\Omega^{2}$ can be computed from \eq{3.6} for a given $\lambda$-order $n$ of $f(R)$ as follows:
\be\n{3.6a}
\Omega^{2}=1+\sum_{k=1}^{n}\lambda^{k}\left(2P_{k}+\sum_{l=1}^{k-1}P_{k-l}P_{l}\right)\,,
\ee
where for brevity we dropped the $[R]$ notation in the $P_{k}$'s terms. Substituting \eq{3.6} into \eq{3.5} and using \eq{3.1} we derive the following $\lambda^{n}$ order term:
\be\n{3.7}
\sum^{n-1}_{k=1}(6P_{n-k}\nabla^{2}P_{k}-R\,P_{n-k}P_{k})-2R\,P_{n}+c_{n}R^{n+1}\,,
\ee
where $n\geq1$. Each $\lambda$-order expression should vanish independently modulo the boundary term. As a result, we can find the following nonlinear recurrence relation of order $n-2$ for $P_{n}$:
\be\n{3.8}
P_{n}=\frac{c_{n}}{2}R^{n}+\sum^{n-1}_{k=1}\left(3R^{-1}P_{n-k}\nabla^{2}P_{k}-\frac{1}{2}P_{k}P_{n-k}\right)\,.
\ee
As an example, for $f(R)$ of $\lambda$-order 2 terms are
\be\n{3.9}
P_{1}=\frac{c_{1}}{2}R\hhh P_{2}=\frac{c_{2}}{2}R^{2}+\frac{c_{1}^{2}}{4}\left(3\nabla^{2}R-\frac{1}{2}R^{2}\right)\,.
\ee
They define the following conformal factor:
\be\n{3.10}
\Omega^{2}=1+\lambda^{1}c_{1}R+\lambda^{2}\left(c_{2}R^{2}+\frac{3}{2}c_{1}^{2}\nabla^{2}R\right)\,.
\ee
The recurrence relation \eq{3.8} together with \eq{3.6a} solves equation \eq{3.5} and proves by induction that the perturbative $f(R)$ theory \eq{3.1} is equivalent to the Einstein gravity \eq{2.1}.

\section{Equivalence with Brans-Dicke theory with $\omega_{0}=0$}
	
The Brans-Dicke theory with $\omega_{0}=0$ parameter without matter fields reads (see, e.g. \cite{Sotiriou_2006,RevModPhys.82.451,PhysRevLett.29.137})
\be\n{4.1}
S=\frac{1}{16\pi}\int d^4x\sqrt{-g}\left[\phi\,R-V(\phi)\right]\,,
\ee
where $\phi$ is the Brans-Dicke scalar field and $V(\phi)$ is its potential. The corresponding field equations are
\ba
G_{\alpha\beta}&=&\frac{1}{\phi}(\nabla_{\alpha}\nabla_{\beta}\,\phi-g_{\alpha\beta}\nabla^{2}\phi)-\frac{g_{\alpha\beta}}{2\phi}V(\phi)\,,\n{4.2a}\\
R&=&V'(\phi)\,,\n{4.2b}
\ea
where the prime stands for the derivative with respect to $\phi$. Assume now that the scalar field depends perturbatively on $R$ and consider the following expansion:
\be\n{4.3}
\phi(R)=\sum_{k\geq0}a_{k}\lambda^{k}R^{k}\,,
\ee
where $\lambda\ll1$ as before and $a_{k}$'s are constant expansion coefficients. Then, integrating the expression \eq{4.2b} and imposing the condition $V(0)=0$ we derive
\be\n{4.4}
V(R)=\sum_{k\geq0}\frac{a_{k}k}{k+1}\lambda^{k}R^{k+1}\,.
\ee
Thus we see that taking $a_{0}=1$ and $a_{k}=c_{k}(k+1)$ for $k\geq1$ the Brans-Dicke action \eq{4.1} is equivalent to the $f(R)$ theory \eq{2.2} and, as it follows from Sec.~II, can be transformed to the Einstein gravity \eq{2.1}.
	
\section{Equivalence with Einstein-scalar field model}
	
The Brans-Dicke model \eq{4.1} can be viewed as Einstein-scalar field model in the Jordan frame. Here we present it in the Einstein frame. Let us first rewrite the action \eq{4.1} in the ``barred''  form
\be\n{5.1}
S=\frac{1}{16\pi}\int d ^4x\sqrt{-\bar{g}}\left[\phi\,\bar{R}-V(\phi)\right]\,.
\ee
Then, we apply the conformal transformations \eq{3.2}, \eq{3.3} with the conformal factor 
\be\n{5.2}
\Omega^{2}=\phi^{-1}\,.
\ee
and present the kinetic term modulo the boundary term as follows: $6\phi^{1/2}\nabla^{2}\phi^{-1/2}\circeq -6(\nabla\phi^{1/2})(\nabla\phi^{-1/2})$. This brings us to the action
\be\n{5.3}
S=\frac{1}{16\pi}\int d ^4x\sqrt{-g}\left[R-\frac{3}{2}\phi^{-2}(\nabla\phi)^{2}-\phi^{-2}V(\phi)\right]\,.
\ee
The final step is to write the kinetic term in the standard form by defining new scalar field $\varphi$, such that
\be\n{5.4}
\phi=\exp\left(\varphi/\sqrt{3}\right)\,
\ee
and the related potential
\be\n{5.5}
U(\varphi)=\exp\left(-2\varphi/\sqrt{3}\right)V(\varphi)\,.
\ee
This results in the Einstein-scalar field model
\be\n{5.6}
S=\frac{1}{16\pi}\int d ^4x\sqrt{-g}\left[R-\frac{1}{2}(\nabla\varphi)^{2}-U(\varphi)\right]\,.
\ee
This procedure is well known and presented in many works (see, e.g. \cite{Sotiriou_2006,RevModPhys.82.451}).  Our goal is to establish its relation with the perturbative $f(R)$ model \eq{2.2}, \eq{3.1}, and, as a result, with the Einstein gravity \eq{2.1}. We observe that relations inverse to \eq{5.2}, \eq{5.4}, and \eq{5.5} transform the Einstein-scalar field model \eq{5.6} to the Brans-Dicke model \eq{4.1}. Then, as it was illustrated in the previous section, the Brans-Dicke model can also be transformed perturbatively via the expressions \eq{4.3} and \eq{4.4} to the Einstein gravity \eq{2.1}.
	
\section{Equivalence maps}
	
In the previous sections we established off-shell equivalence between perturbative $f(R)$ model, the Einstein gravity, the Brans-Dicke with $\omega_{0}=0$ and the Einstein-scalar field models. Here we explore this equivalence in detail. 
	
It is well known that $f(R)$ gravity is equivalent to Brans-Dicke theories (see, e.g. \cite{Sotiriou_2006,RevModPhys.82.451} and the equivalence map presented therein.) This equivalence exists for any $f(R)$ function of class $C^{2}$ and is established by means of the Legendre and conformal transformations. Equivalence between the Einstein gravity and $f(R)$ model can be established by means of the conformal transformation \eq{3.2} where the conformal factor solves equation \eq{3.5}. This equation is a nonlinear PDE of the first (or ignoring boundary term second) order. We do not know whether a general solution to this equation for arbitrary $f(R)$ of class $C^{2}$ exists and if so, how to find it. It could be that there is some class of transformations or generating techniques allowing to find for a given $f(R)$ corresponding solutions to this equation. An analysis of this problem goes beyond the scope of this work. 
	
We were able to find a perturbative solution to this equation for the corresponding form of $f(R)$ function \eq{3.1}. 
This equivalence implies that the perturbative $f(R)$ and the related to it Brans-Dicke and Einstein scalar field models are conformally equivalent to the Einstein gravity. In other words, all these models lie within the Einstein gravity and can be ``perturbatively revealed" by a suitable conformal factor. One can say accordingly, that the Einstein theory is rich enough to accommodate perturbatively via conformal transformation $f(R)$ and related gravity models.

\textcolor{black}{\section{Scenario in the Presence of matter and applications to cosmology} 
As the current work deals with vacuum scenarios, it is important to note that the scenario maybe completely different in the presence of matter. While it is clear from this work that in vacuum the extra degree of freedom is suppressed infinitely for $ f\left(R\right) $ theories (and related theories that $ f\left(R\right) $ can be mapped to), the presence of matter will lead to complications. To be precise, the Lagrangian for the matter fields contain the metric; any changes to the metric, like a conformal transformation, will lead to a change in the matter Lagrangian. In effect, the conformal transformation in the current work and other such transformations will lead to the non-minimal coupling of the matter field with gravity, resulting in an actual deviation from GR. As an example, consider the conformal transformation on the metric which takes an $ f\left(R\right) $ theory in Jordan frame to its equivalent Einstein+scalar model in the Einstein frame. The conformal transformation is given as follows
\begin{eqnarray}
	\tilde{g}_{\mu\nu} &=& f'\left(R\right)\,g_{\mu\nu},
\end{eqnarray}
where $ f'\left(R\right)=\frac{df\left(R\right)}{dR} $. In the Einstein frame, due to this particular conformal transformation, the matter Lagrangian picks up the conformal factor and hence becomes non-minimally coupled to gravity. This particular non-minimal coupling leads to a variable mass of the scalar field which depends on the ambient matter density \cite{Burrage:2017qrf}. To be precise, the mass of the scalar field becomes heavy in regions of high mass density (leading to a sharp fall off of the scalar field profile outside a mass distribution) like the surface of planets, stars, etc. On the other hand when the mass density is low, the mass of the scalar field becomes light, which can support cosmic acceleration. This effect is known as the Chameleon mechanism. Now consider the leading order $ f\left(R\right) $ theory of the form 
\begin{eqnarray}
	f\left(R\right) &\equiv& R\,+\,\alpha\,R^2.
\end{eqnarray}
The conformal transformation required to take this theory into the Einstein frame is then given by
\begin{eqnarray}
	f'\left(R\right) &=& 1\,+\,2\,\alpha\,R.
\end{eqnarray}
Compare this with the leading order conformal transformation for taking the $ \left(R\right) $ theory to Einstein gravity in Eq. (\ref{3.10}) which goes as $ 1\,+\,\lambda\,c_1\,R $. This is not a coincidence. While the former serves to convert higher powers of the Ricci scalar (and in essence higher powers of derivatives of the metric) into kinetic and potential terms of the scalar field, the latter `pushes' the quadratic powers of the Ricci scalar into cubic and higher powers. Both conformal transformations have the same effect of getting rid of the quadratic Ricci term. Both conformal transformations will lead to a non-minimal coupling of the matter fields with gravity, and in return will generate Chameleon screening of the scalar field, which then becomes important for cosmological scenarios.}
	
\section{Discussions and future work} 
In this article we were able to show the pertubative equivalence between polynomial $ f\left(R\right) $ theories of gravity and GR. At the same time equivalences between Brans-Dicke theories, scalar-tensor theories, and polynomial $ f\left(R\right) $ were also reviewed. This exercise shows that it is possible to generate theories of higher complexity and field content from GR itself through proper field redefinitions. Specifically, the graviton, with it's three degrees of freedom in $f\left(R\right)$ theories of gravity is seen to apparently shed one massive scalar degree of freedom through the particular field redefinition in Eqs. (\ref{3.2}) and (\ref{3.10}), and have only two massless tensor degrees of freedom of GR, to which it was mapped to. The equivalences shown imply that if a solution to the vacuum field equations of GR is known, one can generate the corresponding solutions of the vacuum field equations of the modified theories of gravity that were discussed in this article. However, this will require further study and will be published elsewhere.

The equivalences raise the question about the `physicality' of the conformal `frames'. That is, whether a black hole solution in GR in the conformally transformed GR `frame' is the physical solution, or whether the GR frame is simply a convenient frame to generate solutions, and the physical object in question (the black hole solution) is described by the solution of the vacuum field equations of modified gravity frame. However, this question is beyond the scope of this article and will be pursued in a following publication.

In this article, we have only pursued vacuum Lagrangian densities. However, field redefinitions of the form (\ref{3.2}) induce non-minimal coupling of the curvature with the matter Lagrangian density since it is an explicit function of the metric tensor as well. This will lead to the comformally transformed Lagrangian having a more complicated form in the GR `frame' and require further investigation.
	
For future work we will also generalize from $ f\left(R\right) $ theories to the fully general Lagrangian of Eq. (\ref{1.1}), and will aim to come up with a field redefinition that maps it to GR.
	
\begin{acknowledgments}
	The author would like to acknowledge Andrey Shoom for giving significant help and directions throughout the duration of the study. The author would also like to thank Alok Laddha and Ashoke Sen for giving helpful comments. The author wishes to express his gratitude to the MPI f\"ur Gravitationsphysik (Hannover) and ICTS-TIFR (Bangalore) for hospitality. \textcolor{black}{The author would also like the thank the anonymous reviewers for their questions and suggestions.}
\end{acknowledgments}

\appendix
\section{Derivation of the master equation}\label{A}
From Eq. (\ref{3.4}), one has the following
\begin{eqnarray}\label{A.1}
	\sqrt{-g}\,\left[R \,+\, f\left(R\right)\right] \,-\, \sqrt{-\bar{g}}\,\,\bar{R} &=& 0.
\end{eqnarray}
The conformally transformed square root of metric determinant and the Ricci scalar transform as
\begin{eqnarray}
	\sqrt{-\bar{g}} &=& \Omega^4 \, \sqrt{-g} , \label{A.2} \\
	\bar{R} &=& \frac{R}{\Omega^2} \,-\, \frac{6 \, \Box\Omega}{\Omega^3}. \label{A.3}
\end{eqnarray}
Substituting the above in Eq. (\ref{A.1}) leads to the following
\begin{equation}\label{A.4}
	 \sqrt{-g} \left[R\,+\,f\left(R\right)\right]\,-\,\sqrt{-g} \left[\Omega^2\,R\,-\,6\,\Omega\,\Box\Omega\right]\,=\, 0.
\end{equation}
Canceling $ \sqrt{-g} $ on both sides one obtains
\begin{eqnarray}
	6\Omega\Box\Omega-(\Omega^{2}-1)R+f(R)=0.
\end{eqnarray}
\begin{comment}
	Under the transformation
\begin{equation}
	\Omega^2 \,=\, K \label{A.5}
\end{equation}
Eq. (\ref{A.4}) becomes, after some algebra
\begin{eqnarray}
	&&2 \,R \, K \,\left(1 \,-\, K\right) \,-\, 3 \, \nabla_\alpha K \, \nabla^\alpha K + \, 6 \, K \, \Box K \nonumber\\
	&&\,+\, 4 \,m_p^2 \, K \, f\left(R\right) \,=\, 0 \label{A.6}
\end{eqnarray}
Now, using the following identity
\begin{equation}\label{A.7}
	K \, \Box K \,\mathring{=}\, -\nabla_\alpha K \, \nabla^\alpha K ,
\end{equation}
in Eq. (\ref{A.6}), one obtains the master equation
\begin{eqnarray}\label{A.8}
	9\,\nabla_\alpha K\, \nabla^\alpha K\,+\,2\,K\,\left(K\,-\,1\right)\,R\,-\,4\,m_p^2\,K\,f\left(R\right) \mathring{=} 0 \nonumber\\
\end{eqnarray}
\end{comment}

%\bibliography{biblio.bib}
%apsrev4-2.bst 2019-01-14 (MD) hand-edited version of apsrev4-1.bst
%Control: key (0)
%Control: author (8) initials jnrlst
%Control: editor formatted (1) identically to author
%Control: production of article title (0) allowed
%Control: page (0) single
%Control: year (1) truncated
%Control: production of eprint (0) enabled
%

\end{document}